\pdfoutput=1
\documentclass[sigconf,screen=true,anonymous=false,bookmarks=false,9pt]{acmart}
\usepackage{enumitem}
\setlist{leftmargin=5.08mm}

\iftrue
\usepackage{titlesec}
\titlespacing\section{2pt}{10pt plus 1pt minus 1pt}{2pt plus 1pt minus 1pt}
\titlespacing\subsection{2pt}{6pt plus 1pt minus 1pt}{1pt plus 1pt minus 1pt}
\titlespacing\subsubsection{2pt}{5pt plus 1pt minus 1pt}{2pt plus 1pt minus 1pt}
\fi

\usepackage{lipsum}
\usepackage{graphicx}
\usepackage{amsmath}
\usepackage{footnote}
\usepackage{mathtools}
\usepackage{mathrsfs}     
\usepackage{comment}
\usepackage[subrefformat=parens,labelformat=parens]{subfig}
\usepackage{bm}
\usepackage{multirow}
\usepackage{threeparttable,booktabs}
\usepackage{blkarray}
\usepackage{tikz}
\usetikzlibrary{positioning,calc,fit,decorations.pathmorphing,shapes.geometric,shapes.gates.logic.US,calc}
\usepackage{balance}
\usepackage{courier}                                       
\usepackage{cleveref}                                      
\usepackage[mathcal]{eucal}
\usepackage[]{algpseudocode}                               
\algrenewcommand\textproc{\texttt}
\makeatletter\let\float@addtolists\relax\makeatother
\usepackage{algorithm}
\algtext*{EndIf}
\algtext*{EndFor}
\algtext*{EndWhile}

\usepackage{filecontents}                                  
\usepackage{pgfplots}
\pgfplotsset{compat=1.14}
\usepackage{pgfplotstable}
\usepackage{makecell}
\pgfplotsset{compat=newest}
\usepackage[figuresright]{rotating}

\theoremstyle{plain}

\theoremstyle{definition}

\algrenewcommand\textproc{\texttt}

\usepackage[skip=1pt]{caption}            
\definecolor{CUHKorange}{RGB}{244,106,18} 
\definecolor{CUHKblue}{RGB}{0,111,190}    
\definecolor{CUHKgreen}{RGB}{0,127,128}   
\definecolor{CUHKred}{RGB}{228,46,36}     
\definecolor{CUHKyellow}{RGB}{198,148,34} 
\definecolor{CUHKdark}{RGB}{114,44,114}   
\definecolor{CUHKmiddle}{RGB}{144,44,144} 

\usepackage{tcolorbox}
\tcbuselibrary{skins,breakable}
    {\endtcolorbox}

\graphicspath{{./figs/}{../}}

\copyrightyear{2025} 
\acmYear{2025} 
\setcopyright{rightsretained} 
\acmConference[ASPDAC '25]{30th Asia and South Pacific Design Automation
Conference}{January 20--23, 2025}{Tokyo, Japan}
\acmBooktitle{30th Asia and South Pacific Design Automation Conference
(ASPDAC '25), January 20--23, 2025, Tokyo,
Japan}\acmDOI{10.1145/3658617.3703134}
\acmISBN{979-8-4007-0635-6/25/01}

\begin{document}

\title{The Survey of Chiplet-based Integrated Architecture: An EDA perspective}

\iftrue

\author{
    Shixin Chen$^{1}$, \quad
    Hengyuan Zhang$^{2}$, \quad
    Zichao Ling$^{2}$, \quad
    Jianwang Zhai$^{2, \dag}$, \quad
    Bei Yu$^{1, \dag}$\\
    $^{1}$The Chinese University of Hong Kong \\
    $^{2}$Beijing University of Posts and Telecommunications\\
    \{sxchen22, byu\}@cse.cuhk.edu.hk, \, \{zhy679, lingzichao, zhaijw\}@bupt.edu.cn
}

\fi

\renewcommand{\shortauthors}{Shixin Chen, Hengyuan Zhang, Zichao Ling, Jianwang Zhai, and Bei Yu}

\begin{abstract}

Enhancing performance while reducing costs is the fundamental design philosophy of integrated circuits (ICs). 
With advancements in packaging technology, interposer-based chiplet architecture has emerged as a promising solution. 
Chiplet integration, often referred to as 2.5D IC, offers significant benefits, including cost-effectiveness, reusability, and improved performance. 
However, realizing these advantages heavily relies on effective electronic design automation (EDA) processes. 
EDA plays a crucial role in optimizing architecture design, partitioning, combination, physical design, reliability analysis, \textit{etc.} 
Currently, optimizing the automation methodologies for chiplet architecture is a popular focus; therefore, we propose a survey to summarize current methods and discuss future directions.
This paper will review the research literature on design automation methods for chiplet-based architectures, highlighting current challenges and exploring opportunities in 2.5D IC from an EDA perspective. 
We expect this survey will provide valuable insights for the future development of EDA tools chiplet-based integrated architectures.
\end{abstract}

\maketitle

\renewcommand{\thefootnote}{}
\footnotetext{$^{\dag}$Corresponding authors.}

\section{Introduction}

The challenge of designing complex electronic systems that deliver high performance while keeping costs low has been a driving force in integrated circuit (IC) design. 
The concept of Moore's law~\cite{moore1998cramming} has historically guided the industry, illustrating how technological advancements enable the integration of more transistors on a single die, thereby enhancing performance and reducing costs. 
However, chip scaling improvements resulting from the decreasing size of transistors are no longer effective today. 
Firstly, physical limitations make it increasingly difficult to reduce transistor sizes further. 
This results in a bottleneck in improving performance by placing more transistors in a unit area of the silicon wafer. 
Secondly, manufacturing costs continue to rise with the size reduction of transistors. Smaller transistors require more sophisticated and complex lithography systems, which significantly increase manufacturing costs for the chip industry. 
Moreover, the yield from silicon wafers also decreases, making it impossible to maintain cost benefits while improving overall performance.

\begin{figure}[t]
    \includegraphics[width= 0.92\linewidth, bb=0 0 439 229]{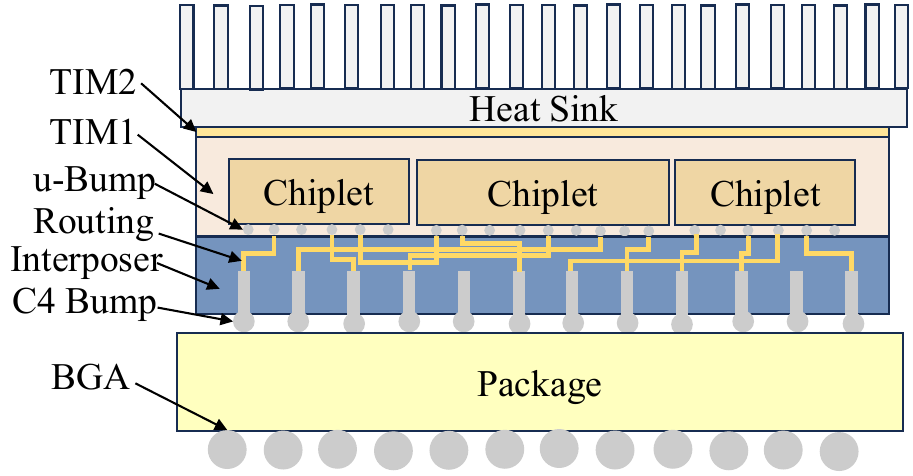}
    \caption{The architecture of chiplet-based 2.5D IC.}
    \label{fig:chiplet-package}
\end{figure}

In response, many leading foundries, such as TSMC, Samsung, and Intel, are actively investigating alternative strategies to lower wafer costs and improve production yields~\cite{zhuang2022multi-package}. 
One promising avenue is the adoption of advanced heterogeneous integration and multi-chiplet architectures. 
This design philosophy involves creating separate hardware modules with specific functions, which are then combined through an interposer to form a comprehensive system~\cite{ECTC2008interposer-source}. 
\Cref{fig:chiplet-package} illustrates the structure of a chiplet-based system fabricated using the advanced chip-on-wafer-on-substrate (CoWoS) technology~\cite{CoWos2017}. 
The interposer-based 2.5D architecture has been used in products such as the Xilinx Virtex-7 2000T FPGA~\cite{eptc2013FPGA-Chiplet} and the AMD ZEN2 Processor~\cite{AMD2020ZEN}.

\begin{figure}[t]
    \centering
    \includegraphics[width=0.96\linewidth, bb=0 0 553 521]{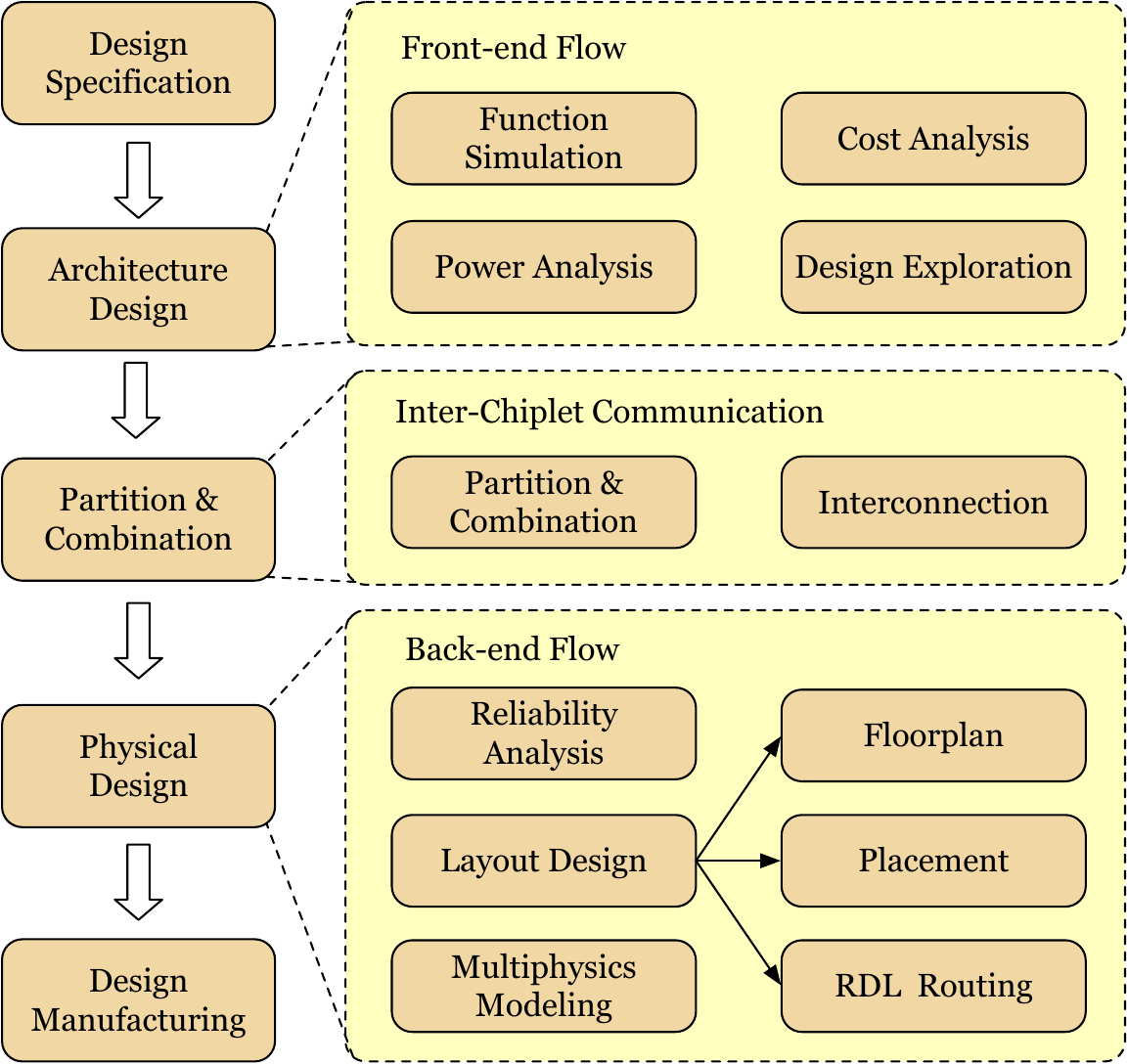}
    \caption{The EDA flow of the chiplet-based architecture.}
    \label{fig:chiplet-eda}
\end{figure}

In the design and implementation of 2.5D architectures, electronic design automation (EDA) plays a crucial role, starting from the front end with architectural design and performance simulation, to the back end with physical design and package design. 
\Cref{fig:chiplet-eda} shows the typical stages in the EDA flow for 2.5D architecture. 
At the design stage, EDA tools facilitate the simulation and exploration of various chiplet configurations, enabling designers to evaluate different architectural choices and their potential impacts on performance, power consumption, cost, \texttt{etc}. 
This early-stage analysis is essential for identifying potential bottlenecks and ensuring that the design meets the desired specifications. 
For physical design and package design, EDA assists in the precise arrangement and interconnection of chiplets on a silicon interposer, accounting for factors such as thermal management and communication latency. 
EDA tools also support reliability analysis and multi-physics modeling to ensure the final assembly is efficient and manufacturable.

This paper reviews the critical role of EDA in the design flow of 2.5D IC and explores future directions to address current challenges. 
Section \Cref{sec:arch-design} highlights the benefits of 2.5D architecture, summarizing research in architectural design, functional simulation, and other issues encountered in the early design stages. 
In Section \Cref{sec:partition-interconnection}, we analyze strategies for partitioning chiplets and their interconnection mechanisms. 
Section \Cref{sec:physical-design} discusses the physical design workflow for 2.5D IC, emphasizing the importance of floorplan \& placement, Routing, multiphysics modeling, and other topics in physics design. 
Next, Section \Cref{sec:future-work} explores current challenges and potential advancements in EDA methodology for 2.5D integration. 
Finally, Section \Cref{sec:conclusion} concludes the paper.

\section{Architecture Design of Chiplet}
\label{sec:arch-design}

\subsection{Chiplet-based 2.5D Architecture}


2.5D integrated architecture is a design paradigm that bridges the gap between traditional 2D and advanced 3D architectures. 
In a 2D architecture, all components of the circuit system are integrated on a single silicon die, known as a system-on-chip (SoC). However, this design can restrict performance, scalability, and yield.
In contrast, 3D architectures~\cite{DAC2000-3DIC-source} stack multiple dies vertically, allowing for greater integration but often introducing complexities related to thermal management and manufacturing challenges. 
The 2.5D approach employs a silicon interposer to place multiple chiplets side by side in a single package. 
In \Cref{fig:chiplet-package}, at the top of the architecture are the chiplets, each serving various functions and bonded to the interposer, where wires connecting different chiplets can be routed. 
The interposer is attached to the package substrate with silicon vias, exposing the pins to peripheral hardware such as SRAM or DRAM.

The 2.5D architecture presents several key advantages over traditional 2D and 3D designs. 
Firstly, it enhances performance by providing high-bandwidth communication between chiplets via the interposer, significantly reducing latency and improving data transfer speeds. 
Secondly, it improves power efficiency, as less critical components can be manufactured using more cost-effective technology nodes, while high-performance components can leverage advanced nodes. 
This strategic separation not only optimizes power consumption but also enhances area and cost efficiency, allowing for better utilization of silicon real estate. 
Moreover, chiplets can be reused and combined to quickly meet a wide variety of demands.
Overall, these factors contribute to a more efficient and powerful integrated circuit design, making 2.5D architecture a compelling choice for modern applications.

\subsection{Emerging Applications of Chiplet}

Apart from mature applications in personal CPU, server CPU, GPU, and FPGA products.
2.5D IC technology is particularly advantageous for applications requiring high performance and efficiency, such as large-scale deep neural network (DNN) accelerators and inference engines for large language models (LLMs). 
For DNN accelerators, the architecture's capacity to facilitate massive parallel processing and rapid data transfer between chiplets is crucial for handling the computational demands of contemporary AI models.
Wang et al.~\cite{aspdac2024chiplet-for-AI} explores the advantages and limitations of 2.5D and 3D heterogeneous integration on representative AI algorithms, such as DNNs, transformers, and graph neural networks.
Ankur et al.~\cite{aspdac2024ciplet-generation} proposes the synthesizer to enable no-human-in-the-loop generation and design space exploration of the chiplets for highly specialized artificial intelligence accelerator.

Similarly, LLM inference benefits from the high bandwidth and low latency afforded by 2.5D designs, enabling quicker processing and enhanced responsiveness. 
Chiplet Cloud~\cite{peng2024chipletcloud} proposes a chiplet-based architecture for LLM supercomputers, and the analysis demonstrates that this architecture offers significant cost benefits for serving large generative language models compared to traditional GPU and Tensor Processor Unit solutions.
Yang et al.~\cite{aspdac2024large-scale-chiplet-computing} discusses how can chiplet bring better cost efficiency and shorter time to market by orchestrating heterogeneous chiplets to support LLM computing workloads.
As the demand for advanced AI applications continues to surge, the adoption of 2.5D architecture is expected to grow, promoting innovation and efficiency across various computational domains.

To conclude, there are several reasons to support 2.5D architecture for various applications.
\textbf{Higher Yield}: The chiplet only needs to support a sub-function of the original large system, which also decreases the area of each chip.
A smaller chip area will lead to a higher yield in manufacturing technology, meaning that given the same silicon wafer budget, the number of chips will be higher.
\textbf{Modular Design:} Since each chiplet has its own function and performance, we can use different combinations of chiplets to create new configurations for various scenarios, like AMD's approach~\cite{AMD2020ZEN}, \textit{i.e.,} more computing chiplets for computation-intensive applications in server clusters, while fewer computing chiplets are used for personal computers.
The interconnection philosophy also makes it possible to use chiplets with different technology nodes, where the core modules utilize more advanced technology and the less critical modules use more mature technology to provide greater cost benefits.
\textbf{Design Efficiency:} A mature chiplet that has been validated can be seen as hardware intellectual property (IP), allowing the design house to use these chiplets to design new products more efficiently.
For example, a Lego-like design method~\cite{HPCA2022Hetero-chiplet} utilizes a heterogeneous chiplet-based architecture to make each chiplet portable to support diverse scenarios, like server CPU, AI processor, and processor CPU.
Such design methods will save costs in functional testing, shorten the time from design to market, and therefore bring more benefits.

\subsection{Simulation Methods of Chiplet Architecture }

Simulation plays a critical role in the design of 2.5D integrated architectures, particularly in the early evaluations of system performance and functional correctness. 
Given the complexity introduced by chiplet architectures, interposers, and communication interfaces, effective simulation is essential for identifying potential issues prior to manufacturing. 
The structural characteristics of chiplets and interposers can lead to additional overhead, such as increased latency and power consumption, which must be accurately modeled to ensure that the integrated system meets its performance specifications. 
The simulations of the 2.5D system allow designers to explore various configurations and assess the impact of different design choices on overall system efficiency.

However, there is currently no customized simulator specifically designed for chiplet-based architectures.
In the research literature, performance-driven chiplet designs~\cite{Patrick2023rapidchiplet,tvlsi2020dse-for-chiplet} often utilize simulation frameworks built for NoC~\cite{BookSim2013,tomacs2016noxim,carlson2011Sniper}. 
Zhi et al.~\cite{NANOCOM2021multi-chip-simu} propose an inter-simulator process communication and synchronization protocol to leverage single-core simulators for modeling the interposer-level network. Additionally, Floorplet~\cite{chen2023floorplet} employs the system-level simulator Gem5~\cite{GEM5-Simulator} to incorporate chiplet latency information introduced by floorplanning, evaluating the overall performance of chiplet-based architectures.

While these tools provide valuable insights, there are still limitations to the simulation of 2.5D systems, which can be categorized into several key areas:
\begin{itemize}
\item  \textit{Accurate Modeling of Chiplet Interactions}: In 2.5D architectures, inter-chip communication is a critical factor, and accurately modeling these interactions is essential for performance evaluation. 
However, many existing simulation tools fail to adequately consider the complex interactions between chiplets, leading to inaccurate performance predictions.
\item  \textit{Integrate Different Technology Nodes}: One of the primary advantages of 2.5D architecture is the ability to integrate chiplets from different technology nodes. 
However, existing simulation tools often lack support for these heterogeneous components, limiting design flexibility and optimization capabilities.
\item \textit{Simulate Complex Communication Protocols}: As the complexity of multi-chip systems increases, simulation tools must be able to model a variety of communication protocols and interfaces. 
However, current tools often have limited capabilities in this regard, which may fail to capture the true dynamics of inter-chip communication. 
\end{itemize}

Developing specialized simulation tools tailored for chiplet performance evaluation requires addressing these critical challenges to enhance the effectiveness and applicability of simulation tools in evaluating 2.5D architectures.

\subsection{Power Analysis of Chiplet Architecture}

Power simulation requires careful consideration of the additional complexity introduced by chiplet-based designs. 
Unlike traditional monolithic architectures, chiplet designs often comprise heterogeneous components that operate and communicate under varying workloads, which complicates power estimation.

To address these challenges, existing research typically employ methods such as static power analysis and dynamic power modeling. 
For instance, McPAT~\cite{McPAT2013} is widely used to model the power of SoCs and has been calibrated using machine learning (ML) to analyze chip power in the advanced technology node~\cite{mcpat-calib2023TCAD}. 
Zhai et al.~\cite{zhai2023tlmodel} use transfer learning to improve the transferability of power models across designs.
PANDA~\cite{zhang2023panda} combines the advantages of analytical and ML power models and further supports power prediction for unknown new technology nodes.
However, these approaches may not fully account for the unique thermal and operational characteristics of individual chiplets or the interposer.
Consequently, there is a pressing need for new methodologies that can accurately assess power consumption across different chiplets and their interactions. 
Recent research has begun to tackle this issue. For example, Kim et al.~\cite{tcpmt2021chiplet-power} present an effective methodology for co-designing the system-level optimization of chiplet power delivery network (PDN) in 2.5D IC designs using commercial EDA tools.
Some AI-based methods are also proposed to manage the power supply system for the chiplet-based architecture.
For example, Li et al.~\cite{ISVLSI2022chiplt-power-RL} utilizes the deep reinforcement learning method to maximum power efficiency for the online control stage. 

Looking ahead, the dynamic nature of workloads in applications such as AI and machine learning necessitates the incorporation of real-time workload models into power simulations. 
This approach would more effectively capture the impact of workload variations on power consumption. By addressing these additional considerations and refining existing tools or developing new ones, designers can achieve more accurate power simulations, ultimately leading to more efficient and reliable chiplet-based architectures.

\begin{figure}[t]
    \centering
    \includegraphics[width=0.96\linewidth, bb= 0 0 393 222]{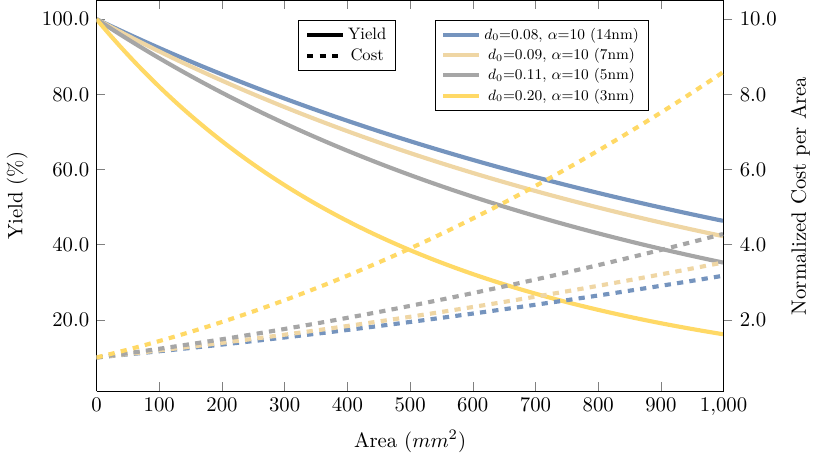}
    \caption{The relationship between cost, yield, and chip area across various manufacturing technology nodes.}
    \label{fig:cost-chiplet}
\end{figure}
\subsection{Cost Analysis of Chiplet Architecture}

The circuit systems incur two main types of costs: non-recurring engineering (NRE) costs and recurring engineering (RE) costs. 
NRE costs are the initial expenses associated with designing a VLSI system, which encompass software development, intellectual property licensing, module or chip design, verification processes, and mask production.
On the other hand, RE costs pertain to the ongoing expenses related to large-scale manufacturing, including wafer fabrication, packaging, and testing.
The basic concept is that if the production quantity is small, the NRE cost is dominant; otherwise, the NRE cost becomes negligible if the quantity is large enough.

In the context of modern architectures, such as chiplets, understanding these cost dynamics is crucial for optimizing design and manufacturing strategies. 
The cost analysis of 2.5D architecture differs from that of traditional monolithic SoCs, with the core idea being to build a specific cost model that considers all factors in design and manufacturing. 
The yield model serves as the foundation for 2.5D architecture, with Seed’s model and the Negative Binomial model being the most widely used~\cite{cunningham1990yield-model}.
\Cref{fig:cost-chiplet} shows the cost and yield relationship between the manufacturing technology node, which further demonstrate that smaller area of chiplet will bring more cost benefit.
Literature~\cite{ahmad2022chiplet-cost-yield,feng2022cost,tang2022cost} propose different cost models from various perspectives. 
Chiplet Actuary~\cite{feng2022cost} presents a quantitative cost model tailored for multi-chip systems, leveraging three representative multi-chip integration technologies. 
Ahmad et al.~\cite{ahmad2022chiplet-cost-yield} introduce an open-source model that provides details on costs—such as material, testing, Know-Good-Die, and operations—for engineering and supply chain personnel to make early trade-off decisions regarding chiplet architecture. 
Additionally, Tang et al.~\cite{tang2022cost} develop an analytical cost model for the 2.5D chiplet system, considering various interconnection options and technology nodes. This study conducts a series of case analyses to explore cost characteristics under both homogeneous and heterogeneous scenarios.

In the cost analysis of chiplet systems, technical complexity presents significant challenges, especially with heterogeneous integration. 
Fortunately, AI technology offers valuable opportunities to enhance cost evaluations. By utilizing historical data, AI can create predictive models for rapid assessments of costs related to various design options, improving decision-making accuracy. 
Additionally, real-time monitoring of cost factors during manufacturing allows for quick identification of potential overruns and optimized resource allocation.
While these advancements offer substantial benefits, challenges regarding system complexity and data quality still exist. 
Overall, integrating AI technology will significantly enhance the efficiency and accuracy of cost evaluations in future chiplet development.

\subsection{Design Exploration of Chiplet Architecture}

Designing efficient 2.5D chiplet systems poses significant challenges due to the vast design space, which differs greatly from traditional monolithic architectures. 
Key challenges arise from the modular and heterogeneous nature of these systems, where factors such as chiplet partitioning, interconnect topology, and packaging must be carefully balanced. 
For example, finer chiplet granularity may improve yield and flexibility, but it can introduce substantial die-to-die communication latency and power overhead. 
The necessity to simultaneously optimize performance, power, cost, and thermal management complicates this process, rendering it both critical and challenging.

To tackle these challenges, various research efforts have introduced tools and methods for design space exploration (DSE). 
DSE methodologies are broadly divided into analytical and machine learning-based approaches.
Analytical methods aim to create lightweight, interpretable models that describe the relationship between hardware design configurations and PPA (performance, power, area), aiding in design space pruning and rapid exploration.
However, these methods often require extensive domain knowledge and face limitations in accuracy and scalability.
To overcome these limitations, a series of studies~\cite{bai2021boom,Zhai2023dse,bai2024towards,Yi2023graphdse,SoCTuner2024ASPDAC,icml2024RL-DSE-multiplier, wang2024circuitgen} have proposed utilizing learning-based approaches to systematically explore chip configurations for microarchitecture design or DNN accelerators. 
These techniques provide scalable solutions for exploring the design space of 2.5D chiplet architectures.

Building on this foundation, Pal et al.~\cite{pal2020DSE-chiplet} developed a framework specifically for the design space exploration of 2.5D architectures, focusing on trade-offs in chiplet partitioning and interconnect strategies to optimize performance and cost. 
Similarly, Iff et al.~\cite{Patrick2023rapidchiplet} introduced RapidChiplet, which enables early evaluations before register-transfer-level design, allowing for quick estimates of power, performance, and cost.

Additionally, workload mapping plays a crucial role in influencing performance and power consumption in 2.5D systems. Cai et al.~\cite{cai2024gemini}proposed the Gemini framework, which integrates layer mapping with architectural adjustments to optimize DNN pipelining and reduce inter-chiplet communication overhead. 
Similarly, Tan et al.~\cite{tan2021nnbaton} introduced NN-Baton, a hierarchical framework that employs the critical-capacity and critical-position method to efficiently manage workload distribution, thereby enhancing energy efficiency.
Collectively, these tools equip researchers and engineers with essential techniques for navigating the complex 2.5D chiplet design space, effectively addressing both architectural and workload-level challenges.

Significant advancements have been made in the DSE of 2.5D chiplet systems, yet many current approaches focus on isolated aspects, such as microarchitecture configurations or workload mapping.
There is a need for an integrated framework that considers multiple design factors—like thermal management, power delivery, and mechanical stress.
Future efforts should aim to create unified DSE methods that incorporate these elements for more accurate design optimization.
As chiplet architectures evolve, integrating AI-driven techniques like reinforcement learning can automate the exploration process and adjust design parameters based on real-time performance data.

\section{Partition and Interconnection of Chiplet}

\label{sec:partition-interconnection}
\begin{figure*}[t]
    \centering
    \includegraphics[width=0.92\linewidth,bb=0 0 1787 587]{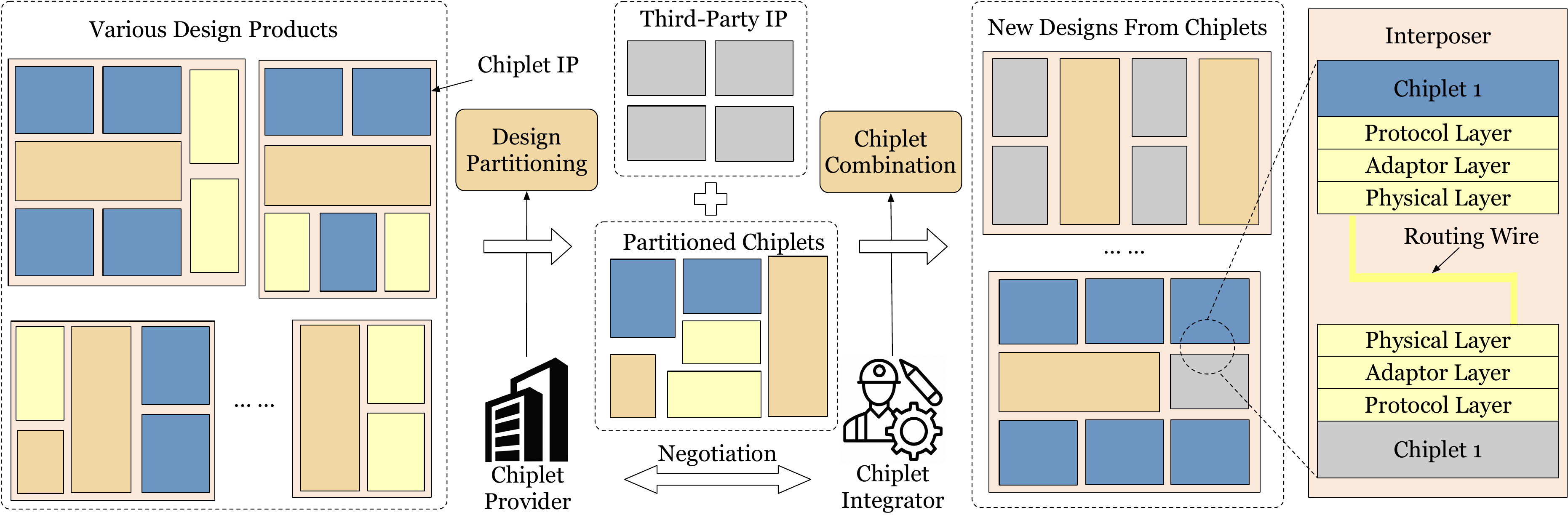}
    \caption{Illustration of partitioning and combining in chiplet-based architecture.}
    \label{fig:partiiton-combination}
\end{figure*}

In 2.5D chiplet systems, partitioning and combining chiplets are critical steps that significantly influence system performance, power consumption, and overall cost. 
\Cref{fig:partiiton-combination} illustrates these processes. 
By utilizing chiplets from existing designs and third-party IPs, designers can create new chip systems with optimized interconnects and communication protocols.

\subsection{{Partition and Combination}}

The primary objective of partitioning is to divide traditional monolithic SoC designs into smaller, independent chiplet modules. 
This modularization enhances flexibility and adaptability to various manufacturing processes, allowing for greater scalability.
However, achieving an optimal partition involves challenges such as managing coupling between functional modules, minimizing communication overhead, and ensuring efficient resource allocation.
Once the system is partitioned, the combination phase focuses on integrating these chiplet modules into a cohesive system. 
This stage requires careful optimization of topology and interconnects to meet performance targets while minimizing costs. Balancing these factors is crucial for the success of a chiplet-based system.

To tackle the challenges of partitioning, various approaches have been developed. 
KaHyPar~\cite{schlag2023kahyper} is a hypergraph partitioning tool that minimizes communication overhead and resource overlap.
Similarly, Chen et al.~\cite{chen2023floorplet} proposed the parChiplet algorithm, which uses a recursive tree structure to optimize the division of SoC modules and packaging layout, addressing communication delays and resource imbalances.
For complex systems, Chipletizer~\cite{Li2024chipletizer} employs multi-layer partitioning and simulated annealing to enhance core reuse and reduce costs. 
Zhuang et al.~\cite{zhuang2022multi-package} introduced a multi-package co-design method that integrates hierarchical programming to distribute dies across multiple packages, improving reliability and scalability.

For the combination, optimizing chiplet selection and interconnect topology is essential for effective communication and system performance. 
Li et al.~\cite{Li2022GIA} proposed GIA that supports adaptive interconnect configurations and provides an automated framework for chiplet selection, topology generation, and layout evaluation, enabling efficient integration of heterogeneous chiplets. 
Pal et al.~\cite{pal2020DSE-chiplet} introduced a framework for exploring chiplet configurations across varied applications, reducing costs while meeting performance constraints and enhancing chiplet reusability.

For network topology, Bharadwaj et al.~\cite{Kite2020DAC} proposed the Kite topology, optimizing chiplet placement to improve throughput, while Patrick et al.’s HexaMesh\cite{HexaMesh2023DAC} adopts a hexagonal layout to minimize communication paths and maximize bandwidth. 
GIA~\cite{Li2022GIA} further includes high-performance interconnects with bypass paths and configurable routers for adaptive communication flows. 
Similarly, Zheng et al.~\cite{zheng2020versatile}proposed the Adapt-NoC framework, which dynamically allocates resources to form tailored sub-networks for specific communication needs. 
Murali et al.~\cite{Murali2006asnoc} enhanced NoC design by integrating floorplan information to enable early timing and power optimization in large networks.

Overall, advancements in partitioning and combining chiplets have significantly improved the adaptability and efficiency of 2.5D systems.
However, increasing heterogeneity presents challenges like managing diverse functionalities and power profiles across chiplets. 
Future research should explore adaptive partitioning methods that dynamically adjust to the specific needs of different chiplets, optimizing the balance between performance, power, and area.
Moreover, as chiplet system complexity grows, integrating thermal management and mechanical stress considerations into the partitioning process will be crucial for stability and reliability. 
Developing automated tools to manage these factors early in the design phase will be vital for seamless integration. 
Addressing these challenges will enhance the flexibility of future 2.5D systems, meeting diverse application needs with improved efficiency.

\subsection{{Interconnection and Communication}}

Effective communication is fundamental to the success of 2.5D chiplet systems, where the modular architecture introduces unique challenges in maintaining system performance, scalability, and power efficiency. 
As chiplets replace traditional monolithic designs, ensuring seamless data flow between these components becomes increasingly complex. 
These systems must support high data transfer rates, low latency, and dynamic routing while managing heterogeneous chiplets that differ in performance, power requirements, and communication protocols. 
The need for high-speed, low-latency communication across chiplets further complicates system integration and configuration, making the design of interconnects and protocols a critical factor in system efficiency.

To address these challenges, various standardized communication protocols have been developed. 
UCIe (unified chiplet interconnect express)~\cite{sharma2022UCIe} simplifies the integration of diverse chiplets by offering a unified interface that supports a wide range of chip types, ensuring compatibility across different manufacturing processes and technology nodes. 
USR (ultra-short reach)~\cite{SSCM2019USR-LINK} and AIB (advanced interface bus) are two additional protocols that enable high-speed die-to-die communication, providing low-latency links that are essential for efficient data transfers between chiplets. 
Moreover, protocols like BoW (bunch-of-wires)~\cite{HOTI2020BoW} offer flexible packaging solutions that enhance communication between chiplets in different systems, supporting various application-specific needs.

Recent innovations have introduced further advancements in the interconnect space based on these protocols. 
For example, Feng et al.~\cite{Feng2023heterogeneousd2d} introduced a heterogeneous interface architecture that integrates parallel and serial interfaces, allowing for flexible communication between chiplets with varying requirements. 
This architecture is particularly advantageous in systems with mixed workloads, where the ability to dynamically switch between communication modes can significantly improve performance and efficiency. 
Additionally, optical communication technologies can offer significantly higher data transfer rates compared to traditional electrical interconnects, enabling large-scale multi-chip systems to handle higher bandwidth demands~\cite{date2020popstar}. 
Li et al.~\cite{Li2024HPPI} proposed HPPI, a reconfigurable photonic interconnect optimized for deep learning accelerators, which enhances dataflow efficiency and reduces energy consumption.

Communication routing algorithms have also evolved to handle the complexities of dynamic traffic and heterogeneous workloads.
Feng et al.~\cite{feng2023scalable} developed the Minus-First Routing algorithm, which combines secure and non-secure flow control strategies to optimize bandwidth utilization and reduce latency.
Similarly, DeFT~\cite{Taheri2022deft} offers a deadlock-free and fault-tolerant routing approach that employs virtual-network separation and dynamic vertical link selection to ensure reliable communication in 2.5D systems.

Despite recent advances, there remain challenges in optimizing communication pathways to accommodate diverse workloads and traffic patterns. Future research should focus on adaptive communication protocols capable of real-time adjustments, ensuring system responsiveness and minimizing latency.
With expanding system scales, addressing power management and maintaining data integrity across communication layers will become increasingly important.
Real-time analytics and AI-driven methods will offer promising ways to manage these complexities, enabling dynamic optimization of communication strategies.



\section{Physical Design of Chiplet-based Architecture}
\label{sec:physical-design}

\subsection{Physical Design Flow}

The physical design flow of 2.5D integration encompasses several critical steps, including floorplanning, placement, routing, multi-physics modeling, and reliability analysis.
\Cref{fig:chiplet-eda} demonstrates the EDA flow for physical designs.

Initially, during the floorplanning stage, designers establish the preliminary layout of various functional blocks on the chip and the interposer, optimizing spatial utilization and performance.
For instance, in the design of high-performance computing systems, careful floorplanning can significantly reduce the distances between processing units, thereby minimizing latency.
Following floorplanning, the placement phase involves the precise positioning of these functional blocks to achieve optimal data transfer rates and minimal delays.
The routing stage is then executed, which involves connecting the signal lines between the various modules to ensure rapid data transmission.
In 2.5D architectures, reliability is critically important due to the complexities of heterogeneous integration.
This integration introduces various factors—such as thermal effects, mechanical stress, and electronic interactions—that can significantly impact overall system performance.

\Cref{tab:chiplet-eda-works} lists the research works that optimize the steps in the physical design flow from an EDA perspective.
While EDA tools have made substantial strides in facilitating 2.5D physical design, further research and innovation are needed to effectively address the increasing complexity and diversity of modern integrated circuit designs.
Recent progress in EDA for 2.5D architecture will be introduced in the following aspects: floorplanning and placement, redistributed layer (RDL) design \& routing strategies, multiphysics modeling, and reliability analysis.


\begin{table}[tb!]
    \centering
    \caption{The Research Works of Physical Design}
    \label{tab:chiplet-eda-works}
    \renewcommand{\arraystretch}{1.0}
    \resizebox{.98 \linewidth}{!}
    {
        \begin{tabular}{ccc}
            \toprule
            Design Stages & Works & Descriptions \\ 
            \midrule
            \multirow{8}{*}{{Floorplan \& Placement}} 
            & \cite{chen2023floorplet} & Performance-aware floorplan \\ 
            & \cite{chiou2023chipletplacement} & Sequence-pair tree placement \\ 
            & \cite{ma2021tap2.5d} & Thermal-aware SA algorithm \\ 
            & \cite{coskun2020TCADcross} & Non-matrix asymmetric SA algorithm \\ 
            & \cite{Yang2022ICCADTransitive} & Transitive closure graph SA algorithm \\ 
            & \cite{duan2024rlplanner} & Chiplet-oriented RL method \\ 
            & \cite{Molter2023bo} & Thermal-aware Bayesian optimization \\ 
            & \cite{deng2024order} & Learning-to-rank based on GNN \\ 
            \midrule
            \multirow{4}{*}{RDL \& Routing} 
            & \cite{kabir2021cross} & Impact of inductive effects on timing \\ 
            & \cite{cai2021simultaneous} & Simultaneous routing and geometry  patterns \\ 
            & \cite{lee2022signal} & Router design in interposer routing \\ 
            & \cite{DACchung2023any} & Any-angle routing for multi-RDL \\ 
            \midrule
            \multirow{4}{*}{Multi-Physics} 
            & \cite{Ansys-Icepak}     & Accurate and time-consuming FEA method             \\
            & \cite{Hotspot7} & Compact thermal model from power trace \\
            & \cite{ma2024electrical} & Models for TSV, bumps, and RDL \\ 
            & \cite{chen2022fast} & Thermal map estimation using GNN \\ 
            \midrule
            \multirow{2}{*}{Other Topics} 
            & \cite{dong2024spiral} & Signal and power integrity analysis \\ 
            & \cite{Miao2024ECTCpdn} & PDN analysis with deep RL \\ 
            \bottomrule
        \end{tabular}
    }
\end{table}

\subsection{Floorplan and Placement}

In this section, we primarily discuss the chiplet floorplan and placement in 2.5D systems. 
In the traditional EDA flow for SoCs, the floorplan provides a high-level layout for the functional blocks and macro cells of the chip, considering their size, orientation, and position, along with factors such as power distribution and thermal management. 
The placement follows floorplanning and determines the specific locations of standard cells or macros within the chip, aiming to optimize connections while satisfying constraints like signal delay, power consumption, and routing resource utilization. 
However, in chiplet-based architectures, placement usually refers to determining the locations of chiplets on the interposer. 
Therefore, in this survey, the terms \textit{placement} and \textit{floorplan} will be used interchangeably.

Regarding the placement of chiplet systems, two key issues arise. 
Firstly, there is a need for an evaluation framework that precisely reflects the impact of the layout design, considering metrics such as thermal performance, communication latency, and warpage. 
Secondly, the layout optimization algorithms must achieve high solution quality and fast solving speed. 
Generally, search algorithms can be classified into three types: \textit{i.e.}, heuristic methods, mathematical analytic optimization, and machine learning approaches.

Heuristic methods, such as enumeration-based algorithms~\cite{liu2014chiplet-floorplan}, branch-and-bound (B\&B) algorithms~\cite{osmolovskyi2018chiplet-placement}, B* Tree~\cite{Floorplan-B-Tree-DAC2000}, sequence pair (SP)~\cite{Floorplan-SP-ICCAD1995}, and corner block list (CBL)~\cite{Floorplan-CBL-ICCAD2000}, have been widely used in the placement for SoC design. 
These methods use a specified data structure to represent the placement solution and utilize the simulated annealing (SA) algorithm to optimize objectives such as area, wirelength, and stress. 
Some works have customized these methods for chiplet-based architectures. 
For example, Chiou et al.~\cite{chiou2023chipletplacement} construct an SP-Tree structure that includes representations of rotated and partial/full sequence pairs and apply the B\&B method to the SP-Tree to seek the optimal layout solution. 
TAP-2.5D~\cite{ma2021tap2.5d} employs a thermal-aware placer based on the SA algorithm, relying on a new placement description data structure to explore the physical design space. 
Coskun et al.~\cite{coskun2020TCADcross} also use the SA algorithm to support arbitrary placements that consider non-matrix and asymmetric chiplet organizations. 
This method benefits the overall 2.5D cost by incorporating a comprehensive micro-bump cost and yield analysis. 
At the same time, Yang et al.~\cite{Yang2022ICCADTransitive} consider warping effects and use a transitive closure graph-based SA algorithm to perturb and optimize these effects more directly.

Another important method for placement optimization is mathematical analytic optimization, which formulates placement or floorplanning as an optimization problem with specific constraints and objectives. 
For example, in traditional placement for SoCs, NTUPlace3~\cite{tcad2008ntuplace3} introduces an analytical placement algorithm that uniquely integrates wirelength, preplaced blocks, and density control through a two-stage smoothing technique and advanced optimization methods. 
In chiplet-based floorplanning, analytic methods are also utilized to handle more complicated constraints, particularly those related to reliability. 
For instance, Chen et al.~\cite{chen2023floorplet} propose optChiplet, a floorplan optimization framework that considers reliability, cost, area, and performance metrics. While this framework accurately accounts for many factors, using the solver to solve the problem can be quite time-consuming (taking several hours for multiple chiplets). 
Zhuang et al.~\cite{zhuang2022multi-package} present a novel approach to solving the multi-package co-design problem using mathematical programming methods, aiming to optimize the trade-offs between inter-package and intra-package costs.
 
With the development of AI, placement algorithms also benefit from ML-based methods~\cite{wang2024placement-bench,duan2024rlplanner,Molter2023bo,deng2024order}. 
For example, Duan et al. introduce RLPlanner~\cite{duan2024rlplanner}, a layout tool based on reinforcement learning suitable for the planning of early chiplet-based systems. 
It utilizes advanced RL technology combined with a novel rapid thermal assessment method. 
Molter et al.~\cite{Molter2023bo} employ Bayesian optimization for thermal-aware chip layout optimization. 
They propose an input space kernel function that includes arrangement and continuous variables, evaluating it using a multi-objective function composed of area, temperature, and temperature gradient, through finite element method thermal simulations. 
Deng et al.~\cite{deng2024order} propose a chiplet placement order ranking framework based on graph neural networks, utilizing the learning-to-rank method to select the optimal chiplet placement order. 
A network architecture primarily based on RankNet~\cite{icml2005RankNet} was designed, combining graph neural networks for feature extraction.

Overall, the existing methods can handle small-scale 2.5D chiplet sets in accept time with good performance.  
However, the scale of future 2.5D chiplet systems may increase to dozens or even more small chips, and the algorithms mentioned above require several hours to process systems containing more small chips. 
Therefore, finding more efficient chip layout algorithms and tools to shorten optimization time is crucial research direction in the physical design of large-scale 2.5D IC.

\subsection{RDL and Routing}

The interposer, also known as a redistributed layer (RDL), is a crucial component in the physical design of chiplets. Its design and the associated wiring issues are important topics of study. The wires on the interposer facilitate communication between chiplets and provide interfaces for peripheral devices. However, the routing of these wires introduces additional latency, and parasitic parameters can significantly affect overall performance and reliability.
In the advanced packaging technologies currently promoted by the industry, such as CoWoS, an intermediary interposer design method is commonly employed. 
This intermediary layer uses wiring within the interposer and through-silicon-via (TSV) technology to connect chiplets and establish connections to the packaging substrate. 

Researchers are increasingly focusing on large-scale redistributed layer design and the associated interconnection wiring challenges. 
For example, Kabir et al.~\cite{kabir2021cross} propose a new collaborative optimization process that takes into account the inductive effects of redistributed layer wires on timing during the design of 2.5D systems.
They also create an RLC delay model based on the properties of redistributed layer interconnections to analyze input/output drivers.
Cai et al.~\cite{cai2021simultaneous} develop an innovative simultaneous routing framework that employs a chord-based tile model and a net-sequence list to search for global routing guides. 
Their approach includes both free-assignment and pre-assignment on the same routing graph. 
Additionally, they introduce a geometry-based pattern routing algorithm to find routing solutions, followed by a net padding method to address any failed routes.
Lee et al.~\cite{lee2022signal} introduce a router designed for interposer routing, which optimizes the integrity of high-speed signal transmission by protecting signal lines and adjusting their width and spacing.
This router adheres to complex design rules to maximize signal integrity and interconnect reliability.
Chung et al.~\cite{DACchung2023any} tackle the challenge of any-angle routing for multi-redistributed layer systems.
They develop an optimization routing algorithm and propose a method for generating routing guides after dynamic triangulation, extending the access point adjustment technique to support multiple lines.

As the chiplet architecture continues to evolve, routing tools will play a pivotal role in addressing both opportunities and challenges in this field. 
The increasing complexity of interconnections among numerous chiplets demands advanced routing techniques that can efficiently manage the associated latency and performance issues. 
Future routing tools must not only optimize wire routing but also incorporate machine learning algorithms to predict and mitigate potential reliability concerns, such as signal integrity and thermal effects.
The development of robust, scalable routing solutions that can accommodate the unique requirements of chiplet-based systems will be crucial for realizing the full potential of this architecture

\subsection{Multiphysics Simulation}

The complexity of the 2.5D package can introduce significant thermal stress risks due to varying material properties, particularly in chiplet technology applications where high-power electromagnetic pulses and self-heating effects are prominent.
The multiphysics effects in advanced integrated packaging are primarily described by the continuity equation for current, heat conduction equations, and elasticity equations. 
Factors such as Joule heating can elevate temperatures, causing thermal deformation and stress, which can alter simulation mesh and affect electric and thermal field distributions. 
Consequently, multiphysics simulation is crucial for analyzing the interactions among electrical, thermal, and mechanical effects. 
By integrating these simulations, it is possible to predict and optimize chip performance in complex packaging environments. 

In 2.5D chip design, various multiphysics simulation tools are widely used to address complex design challenges and multiple physical effects, often employing time-consuming finite element analysis (FEA) methods. For example, Ansys Icepak~\cite{Ansys-Icepak} focuses on thermal management, helping engineers optimize heat dissipation by accurately simulating heat flow and temperature distribution. 
COMSOL Multiphysics provides coupled simulation capabilities for multiple physical fields, enabling users to analyze electrical, thermal, and mechanical effects simultaneously \cite{COMSOL-Multiphysics}.
However, these commercial tools also have some shortcomings, including high costs, steep learning curves, and limitations in accuracy under extreme conditions.

In contrast to commercial tools, researchers tend to develop fast, relatively accurate, and lightweight approaches to reduce computation time and quickly predict metrics to accelerate the development process. HotSpot~\cite{Hotspot7} is a simulation tool used in the early stages of 2.5D architecture design to model thermal effects using the compact thermal model (CTM).
It predicts temperature distributions by analyzing power consumption and heat dissipation, allowing for the creation of detailed thermal models that incorporate various components and environmental factors. 
Ma et al.~\cite{ma2024electrical} present an electrical-thermal coupling model that simultaneously predicts temperature distribution and voltage drop in 2.5D chip heterogeneous integration systems, taking into account the Joule heating effect. 
Its advantages lie in high accuracy and practicality, making it useful for optimizing thermal management and electrical performance in packaging structures.

For future direction, machine learning-based methods are being utilized in multiphysics simulation to speed up the simulation. 
For example, Chen et al.~\cite{chen2022fast} present a novel graph convolutional network architecture for estimating thermal maps in 2.5D chiplet-based systems, leveraging global power features and advanced techniques such as skip connections and edge-based attention.

\subsection{Other Topics}

In addition to the physical design considerations discussed, there are several other critical topics that influence the overall performance and reliability of chiplet systems. 

One such topic is Signal Integrity (SI) and Power Integrity (PI), which plays a vital role in electronic design, particularly concerning the quality and reliability of signal transmission in high-speed digital circuits. 
In chiplet systems, the complexity of interconnect channels, compounded by varying operating frequencies and layouts, makes signal integrity issues increasingly significant.

To address these challenges, the SPIRAL framework~\cite{dong2024spiral} has been introduced for the integrated analysis of SI and PI in high-speed chiplet interconnects.
By employing impulse response models and leveraging machine learning techniques, SPIRAL achieves notable enhancements in processing speed while ensuring low error rates.

Additionally, the design of the power distribution network (PDN) within chiplet systems is another essential consideration. 
Miao et al.~\cite{Miao2024ECTCpdn} present a framework utilizing deep reinforcement learning to optimize the PDN design for 2.5D packaged multi-chiplet systems.
This innovative approach notably reduces computation time and uncovers superior solutions within a vast design space, effectively mitigating issues related to lengthy manual design processes and the scarcity of open-source data.

\section{Challenges and Future Directions}
\label{sec:future-work}


Currently, most commercial EDA tools designed specifically for single-die applications do not adequately support chiplet-based architectures. The entire design process of 2.5D IC is quite complex, with each step presenting new demands and challenges for EDA tools.
In a heterogeneous integration system of 2.5D IC, the function and reliability of circuits are influenced by various factors, including electrical, magnetic, thermal, and mechanical stresses. 
There is currently no comprehensive platform that considers all relevant factors; instead, many single-point tools that cannot provide a holistic view or integrated feedback, potentially reducing design efficiency.
Moreover, early-stage multi-physics modeling tools rely on numerous mathematical and physical models, and the computation-intensive nature of this modeling can slow down the design process, hindering overall design efficiency.


From the perspective of placement and routing, traditional EDA tools often struggle to manage the increased complexity introduced by chiplets. 
The interactions between multiple chiplets necessitate advanced algorithms that can optimize not only for distance but also for signal integrity and power distribution. 
Current tools typically lack the capability to effectively model these inter-chiplet connections, leading to suboptimal layouts.

Additionally, design rule checks for 2.5D IC require more sophisticated approaches, as the rules governing interactions between various chiplets can differ significantly from those of single-die designs. 
Existing tools may not be equipped to handle the nuances of multi-chiplet interactions, increasing the risk of design errors that could compromise performance and manufacturability.

Moreover, the verification processes for 2.5D IC face unique challenges. The traditional methodologies for functional verification may not suffice, as they often overlook the complexities of chiplet communications and shared resources. 
This gap can lead to undetected issues that only manifest during later stages of production, thus increasing costs and time-to-market.

To address these challenges, it is essential to develop EDA tools that support multi-domain simulations and provide integrated solutions that align with the unique requirements of 2.5D IC designs. 
This includes enhancing the interoperability of tools, allowing seamless data exchange, and implementing machine learning algorithms to optimize the design workflow. 
Additionally, the incorporation of real-time feedback mechanisms during the design phase could significantly improve reliability and performance, ultimately leading to more efficient and robust 2.5D IC solutions.

\section{Conclusion}
\label{sec:conclusion}

In this article, we explored the application and advantages of chiplets in 2.5D integrated architectures from an EDA perspective. 
We highlighted the improvements in performance, power consumption, and area efficiency that 2.5D architectures offer compared to traditional 2D and 3D designs, emphasizing the importance of simulation, physical design, communication mechanisms, and modeling in the design process.
By effectively leveraging EDA workflows, design teams can better tackle the complexities involved and achieve efficient 2.5D integrated solutions. 
Looking ahead, with the advancement of technologies such as AI and machine learning, 2.5D architectures are poised to demonstrate even greater potential in high-performance computing and other fields.

{
\section*{Acknowledgment}
This work is supported by the National Key R\&D Program of China (2022YFB2901100), 
the National Natural Science Foundation of China (No.~62404021), 
the Beijing Natural Science Foundation (No.~4244107, QY24216),
the MIND project (MINDXZ202404),
and AI Chip Center for Emerging Smart Systems (ACCESS), Hong Kong.
}





\end{document}